# SEARCH-BASED SOFTWARE TEST DATA GENERATION USING EVOLUTIONARY COMPUTATION

P. Maragathavalli [1]

[1]Department of Information Technology, Pondicherry Engineering College, Puducherry
marapriya@pec.edu

## ABSTRACT

*Search-based Software Engineering has been utilized for a number of software engineering activities. One area where Search-Based Software Engineering has seen much application is test data generation.*

*Evolutionary testing designates the use of metaheuristic search methods for test case generation. The search space is the input domain of the test object, with each individual or potential solution, being an encoded set of inputs to that test object. The fitness function is tailored to find test data for the type of test that is being undertaken.*

*Evolutionary Testing (ET) uses optimizing search techniques such as evolutionary algorithms to generate test data. The effectiveness of GA-based testing system is compared with a Random testing system. For simple programs both testing systems work fine, but as the complexity of the program or the complexity of input domain grows, GA-based testing system significantly outperforms Random testing.*

## 1. INTRODUCTION

Search based optimization techniques have been applied to a number of software engineering activities[1] such as requirements engineering, project planning and cost estimation through testing, to automated maintenance, service-oriented software engineering, compiler optimization and quality assessment. However the above listed applications can be the optimization can be applied over the software engineering activity.

A wide range of different optimization and search techniques have been used. The most widely used methods are local search, simulated annealing, genetic algorithms and genetic programming. However, whatever may be the search technique employed, it is the fitness function that plays a major role and it captures a test objective and makes a contribution to the test adequacy criterion. Using the fitness function as a guide, the search seeks test inputs that maximize the achievement of this test objective.

***Search Based Software Testing (SBST):***

SBST research has attracted much attention in recent years[2] as part of a general interest in search based software engineering approaches. The growing interest in search based software testing can be attributed to the fact that there is a need for automatic generation of test data, since it is well known that exhaustive testing is infeasible and the fact that software test data generation is considered NP-hard problem [3].

The reminder of this paper is organized as follows: Section 2 describes the evolutionary testing. Section 3 describes the optimization techniques including the meta-heuristic search techniques. Sections 4 and 5 comprises of analysis and discussion of results, while the paper is concluded in Section 6.

## 2. EVOLUTIONARY TESTING

Evolutionary testing makes use of meta-heuristic search techniques for test case generation. Evolutionary Testing is a sub-field of Search Based Testing in which Evolutionary Algorithms are used to guide the search. The Fig.1 shows the structure and interaction of test activities including test case design by means of evolutionary algorithms.

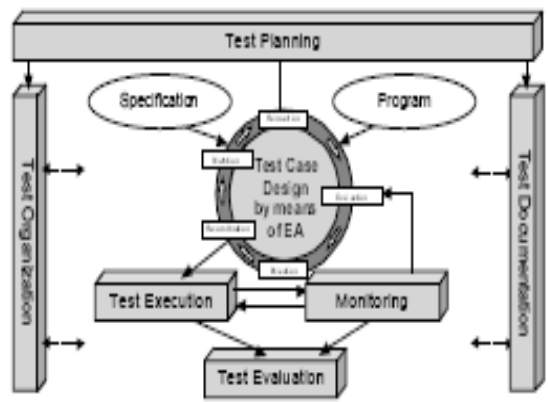

Figure 1. Structure and interaction of test activities including test case design by means of evolutionary algorithms.

The test aim is transformed into an optimization problem. The input domain of the test object forms the search space. The test object searches for test data that fulfils the respective test aim in the search space. A numeric representation of the test aim is necessary for this search. This numeric representation is used to define objective functions suitable for the evaluation of the generated test data. Depending on the test aim pursued, different heuristic functions emerge for test data evaluation.

Due to the non-linearity of software (if-statements, loops etc.) the conversion of test aim to optimization problems mostly leads to complex, discontinuous, and non-linear search spaces. Therefore neighborhood search methods (such as hill climbing), are not recommended. Instead, meta-heuristic search methods are employed, e.g. evolutionary algorithms, simulated annealing or tabu search. Evolutionary Algorithms have proved a powerful optimization algorithm for the successful solution of software testing.

## 3. OPTIMIZATION TECHNIQUES

Some of the optimization techniques that have been successfully[4] applied to test data generation are Hill Climbing(HC) ,Simulated Annealing(SA), Genetic Algorithms(GAs), Tabu Search(TS),Ant Colony Optimization(ACO), Artificial Immune System(AIS), Estimation of Distribution Algorithms(EDAs), Scatter Search(SS) and Evolutionary Strategies(ESs).

### 3.1. Meta-heuristic Search Techniques

Meta-heuristic techniques have also been applied to testing problems in a field known as Search Based Software Testing [2], [3], a sub-area of Search Based Software Engineering (SBSE) [1]. Evolutionary algorithms are one of the most popular meta-heuristic search algorithms and are widely used to solve a variety of problems.

The local Search techniques generally used are

i. Hill Climbing
ii. Simulated Annealing
iii. Tabu Search

**Hill Climbing**

In hill climbing, the search proceeds [6] from randomly chosen point by considering the neighbors of the point. Once a neighbor is found to be fitter then this becomes the current point in the search space and the process is repeated. If there is no fitter neighbor, then the search terminates and a maximum has been found (by definition). However, HC is a simple technique which is easy to implement and robust in the software engineering applications of modularization and cost estimation.

**Simulated Annealing**

Simulated annealing is a local search method. It samples the whole domain and improves the solution by recombination in some form. In simulated annealing a value x1, is chosen for the solution, x, and the solution which has the minimal cost (or objective) function, E, is chosen. Cost functions define the relative and desirability of particular solutions. Minimizing the objective function is usually referred to as a cost function; whereas, maximizing is usually referred to as fitness function.

**Tabu Search**

Tabu search is a metaheuristic algorithm that can be used for solving combinatorial optimization problems, such as the travelling salesman problem (TSP). Tabu search uses a local or neighbourhood search procedure to iteratively move from a solution $x$ to a solution $x'$ in the neighbourhood of $x$, until some stopping criterion has been satisfied. To explore regions of the search space that would be left unexplored by the local search procedure (see local optimality), tabu search modifies the neighbourhood structure of each solution as the search progresses.

### 3.2. Evolutionary Search Using Genetic Algorithms

GA forms a method of adaptive search in the sense that they modify the data in order to optimize a fitness function. A search space is defined, and the GAS probe for the global optimum. A GA starts with guesses and attempts to improve the guesses by evolution. A GA will typically have five parts: (1) a representation of a guess called a chromosome, (2) an initial pool of chromosomes, (3) a fitness function, (4) a selection function and (5) a crossover operator and a mutation operator. A chromosome can be a binary string or a more elaborate data structure. The initial pool of chromosomes can be randomly produced or manually created. The fitness function measures the suitability of a chromosome to meet a specified objective: for coverage based ATG, a chromosome is fitter if it corresponds to greater coverage. The selection function decides which chromosomes will participate in the evolution stage of the genetic algorithm made up by the crossover and mutation operators. The crossover operator exchanges genes from two chromosomes and creates two new chromosomes. The mutation operator changes a gene in a chromosome and creates one new chromosome. Fig.2 shows the generic search based test input generation scheme.

### 3.3 Evolutionary Search Using Genetic Programming

Genetic programming results in a program, which gives the solution of a particular problem. The fitness function is defined in terms of how close the program comes to solving the problem. The operators for mutation and mating are defined in terms of the program's abstract syntax tree. Because these operators are applied to trees rather than sequences, their definition is typically less straight forward than those applied to GAs? GP can be used to find fits to software engineering data, such as project estimation data.

In order to apply metaheuristics [6] to software engineering problems the following steps should therefore be considered:

  i. Ask: Is this a suitable problem?
     That is, "is the search space sufficiently large to make exhaustive search impractical?"
 ii. Define a representation for the possible solutions.
iii. Define the fitness function.
 iv. Select an appropriate metaheuristic technique for the problem.
  v. Start with the simple local search and consider other genetic approaches.

### 3.4 Coverage Criteria

The testing requirements satisfied by the generated test data is the measurement of coverage in terms of statement, condition, path, branch, decision etc.

### 3.4.1 Statement coverage

Statement coverage measures the number of executable statements in the code that are executed by a test suite. 100% statement coverage is achieved when every statement in the code is executed.

### 3.4.2 Decision coverage

Decision coverage, also known as branch coverage, measures the extent to which all outcomes of branch statements (such as if, do-while or switch statements) are covered by test cases.

To achieve decision coverage, two test data $I1$ and $I2$ need to be generated corresponding to each decision $di$ in the program such that $di$ evaluates to true when the code is executed with input $I1$ and evaluates to false when code is executed with input $I2$. For example, to cover the decision at line 70 in Fig.2, we require two test data such that the 'if' condition evaluates to true in one case and false in the other.

```
10: int inp1, inp2; //Inputs
20: int test()
30: {
40: int lVar =0, retVal = 0;
50: if ( inp1 > 15 )
60: lVar = 1;
70: if ( lVar && inp2 )
80: retVal = 1;
90: return retVal;
100: }
```

Figure 2. Sample C code

### 3.4.3. Condition coverage

Condition coverage is similar to decision coverage with the only difference being that for condition coverage, two test data $I1$ and $I2$ are needed for each condition in a decision.

### 3.5. Automated test data generation (ATDG)

Most of the work on Software Testing has concerned the problem of generating inputs that provide a test suite that meets a test adequacy criterion. The schematic representation is presented in Fig.3. Often this problem of generating test inputs is called 'Automated Test Data Generation (ATDG)' though, strictly speaking, without an oracle, only the input is generated. Fig.3 illustrates the generic form of the most common approach in the literature, in which test inputs are generated according to a test adequacy criteria [6]. The test adequacy criterion is the human input to the process. It determines the goal of testing.

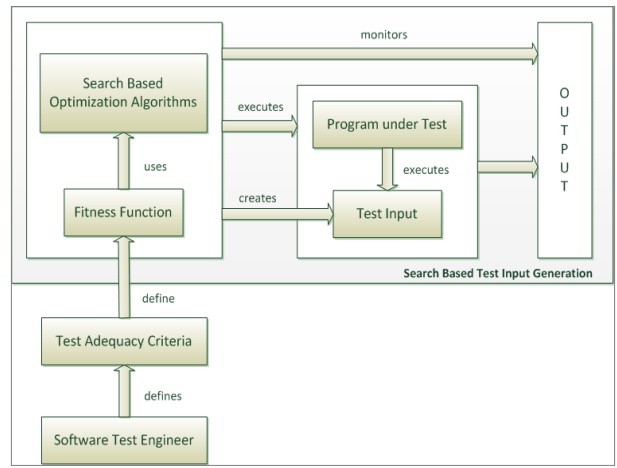

Figure 3. A generic search-based test input generation scheme

The adequacy criteria can be almost any form of testing goal that can be defined and assessed numerically. For instance, it can be structural (cover branches, paths, statements) functional (cover scenarios), temporal (find worst/best case execution times) etc. This generic nature of Search-Based Testing (SBT) has been a considerable advantage and has been one of the reasons why many authors have been able to adapt the SBT approach different formulations.

The fitness function captures the crucial information and it differentiates the good solution from the poor one. Once a fitness function has been defined for a test adequacy criterion, then the generation of adequate test inputs can be automated using SBSE. The SBSE tools that implement different forms of testing all follow the broad structure outlined in Fig.3.

They code the adequacy as fitness, using it to assess the fitness of candidate test inputs. In order to assign fitness to the test inputs, the ATDG system has to make the program to be executed for the inputs. The ATDG system then monitors the execution to assess fitness based on how well the inputs meet the test adequacy criterion.

### 3.6. Existing Applications of Optimization Techniques to Software Engineering Problems

A. Optimizing the search for accurate cost estimates.
B. Optimizing the search for resource allocations in project planning.
C. Optimizing the search for requirements form the next release.
D. Optimizing design decisions.
E. Optimizing source code.
F. Optimizing test data generation
   (structural, functional, non- functional, safety, robustness, stress, mutation, integration and exception).

G. Optimizing test data selection & prioritization.
H. Optimizing module clustering.
I. Optimizing maintenance and reverse engineering.

## 4. ANALYSIS OF THE EXISTING TEST DATA GENERATION TECHNIQUES

The comparative study on the existing test data generation techniques are given in the form of a tabular column [Table 1].

Table 1: Comparative study on the existing test data generation techniques

| SI. No. | Title | Year | Publication | Search techniques | Data set | Parameters considered | Performance metrics | Future directions |
|---|---|---|---|---|---|---|---|---|
| 1. | "Evolutionary white-box software test with the EvoTest Framework, a progress report" | 2010 | *Proceedings of the International Conference on Software Testing, Verification and Validation Workshop* | **Evolutionary Testing** Genetic Algorithm | Java Test Samples, Objective CAML | No. of generated test cases, No. of branches, % of functions | Branch Coverage (%), Test effort | Support for pointers-to-void, array-type, input variables; Improve the compatibility; Evaluation of the tool with a C compiler; |
| 2. | "An Algorithm for Efficient Assertions-Based Test Data Generation" | 2010 | *Proceedings of the International Multi-conference of Engineers and Computer Scientists* | New Heuristic Approach | Existing samples | Array size, longest path length, code coverage | Accuracy, Scalability | To improve the accuracy; |
| 3. | "Breeding Software Test Data with Genetic-Particle Swarm Mixed Algorithme" | 2010 | *Journal of Computers; Vol.5,No.2, FEBRUARY 2010* | GPSMA (Genetic-Particle Swarm Mixed Algorithm) | Benchmark programs - Triangle classification | Size, String length, Probability of crossover, No. of iteration | Efficiency, Convergence Speed, (diversity) | Testing the GPSMA; Applications to R-W optimization problems; Fine-tuning of the parameters; |

| S. No. | Title | Year | Publication | Search techniques | Data set | Parameters considered | Performance metrics | Future directions |
|---|---|---|---|---|---|---|---|---|
| 4. | "Optimizing for the Number of Tests Generated in Search Based Test Data Generation with an Application to the Oracle Cost Problem" | *2010* | *Third International Conference on Software Testing, Verification, and Validation Workshops* | Memory-Based Approach, Greedy Approach, Genetic Algorithm | Empirical Results | No. of test cases generated, Domain size, Cost, Branch Coverage | Complexity, Effectiveness, Diversity | Hybrid algorithm which may be capable of combining the features of both CDG and set cover approaches. |
| 5. | "Automated GUI Test Coverage Analysis using GA" | *2010* | *7-th International conference on Information Technology* | Genetic Algorithm | | Test-path, length, No. of generations, Coverage achieved | Accuracy of fitness, Effectiveness | Developing an automated test generation tool for supporting their approach; Use other optimization techniques. |
| 6. | "Scatter Search" | *2009* | *Information and Software Technology* | Scatter Search - Evolutionary Method | Benchmarks | Range of input variables, No. of test cases generated, % of branch coverage | Efficiency, Diversity | Benchmarks |
| 7. | "Automatic Test Data Generation for C Programs" | *2009* | *3-rd IEEE International Conference on Secure Software Integration and Reliability Improvement* | BLAST – software model checker(new algorithm), SAL – framework | Automotive applications(Embedded domain) | Time, condition coverage | Efforts required | Enhance the tool in order to satisfy more criteria such as boundary value analysis, equivalence partitioning, def-use analysis and to find out possible errors in code |

| S. No. | Title | Year | Publication | Search techniques | Data set | Parameters considered | Performance metrics | Future directions |
|---|---|---|---|---|---|---|---|---|
| 8. | "An Approach to Generate Software Test Data for a Specific Path Automatically with Genetic Algorithm" | *2009* | *8th IEEE International Conference on Reliability, Maintainability and Safety* | Genetic Algorithm | Testing Benchmark-TriType | No. of test data generated, Convergence ability, Consumed time(for searching) | Efficiency, Ability | To decrease the evolutionary status and to improve the search efficiency; To develop completed framework for automated testing data generation |
| 9. | "Test Data Generation Using Annealing Immune Genetic Algorithm" | *2009* | *5-th International Joint Conference on INC, IMS and IDC* | Genetic Algorithm, Simulated annealing algorithm, Immune Genetic Algorithm | Existing Programs-Triangle Classification | Data coverage, no. of generations | Efficiency | Practical usage of AIGA for software testing |
| 10. | "Comparison of Two Fitness Functions for GA-based Path-Oriented Test Data Generation" | *2009* | *5-th International Conference on Natural Computation* | Genetic Algorithm (i)Branch distance (ii)Normalized extended Hamming distance | Sample program – Triangle Classification | Length of chromosome, Test data coverage | Efficiency | Experiment on larger and more complex programs |
| 11. | "Generation of Test Data Using Meta Heuristic Approach" | *2008* | Proceedings of IEEE Region 10 Conference on TENCON 2008 | GA, Ant Colony Optimization algorithm-Resource request algorithm | Existing test samples | Success ratio | Effectiveness | To increase the success ratio |
| 12. | "Using a Genetic Algorithm and Formal Concept Analysis to Generate Branch Coverage Test Data Automatically" | *2009* | *4-th International Conference on Computer Science and Education* | **A** new path selection algorithm (using Fibonacci series) | Existing Programs | Statement Coverage, Data flow coverage | Cost, Computation load | Can handle different language procedures |
| 13. | " Automatic Test Data Generation for Multiple Condition and MCDC Coverage" | *2009* | *4th International Conference on Software Engineering Advances* | Search technique used : **Simulated Annealing;** Proposed framework [tool] – based on FF calculation | Benchmark programs – [Triangle, Quadratic …] Open source **API JGAP** java genetic algorithms package | FF, CF, Success rate, Coverage, Avg. time of execution | 'Control flow coverage', Multiple condition Decision coverage | Tool can be extended to include other search-based algorithms; |

# 5. DISCUSSION ON THE VARIOUS SEARCH TECHNIQUES

Almost in all cases, the meta-heuristic search techniques have been implemented for the specific application for e.g., multi-objective NRP, Ajax web applications, triangle classification problem, software clustering problem, project resource allocation, signal generation, buffer overflow problem, network security, safety, R-T tasks & in fault prediction. In most of the combinatorial problems, they have got better results by implementing Evolutionary Algorithms such as GAs, SA, TS, GP and they compared their results with the local search such as RS, HC. The dataset used were taken from various sources. The main quality parameters considered are branch, path coverage, accuracy and fitness. The results obtained in random generation for few sample programs are given in a form of table [2].

Table 2: Results obtained from sample programs

| Program name | Range of the input variables | % of the branch coverage | No. of test cases generated | Time consumed (in secs) |
|---|---|---|---|---|
| Linear search | 1 to 50 | 95.26 | 10 | 1.78 |
| Quadratic equation | -10 to 10 | 83.33 | 5 | 3.66 |
| Bubble sort | 1 to 40 | 89.64 | 8 | 5.93 |
| Triangle classification | 1 to 20 | 81.56 | 5 | 4.01 |
| Greatest common divisor | 1 to 100 | 80.3 | 10 | 1.97 |
| Binary search | 1 to 35 | 80.1 | 8 | 5.83 |

Triangle classification program results using genetic algorithm are as follows:

Population size: 4, 2

Range: [5, 15]

| Generation | A | B | C | Fitness |
|---|---|---|---|---|
| Generation 1 | 14 | 5 | 9 | 2.0 |
| | 14 | 14 | 11 | 10.0 |
| | 14 | 9 | 8 | 10.0 |
| | 12 | 6 | 12 | 11.0 |
| Generation 2 | 14 | 14 | 9 | 10.0 |
| | 14 | 5 | 11 | 11.0 |

Range: [-10, 20]

Comparison of genetic algorithm with random testing is given below:

| Testing | 1 | 2 | 3 | 4 | 5 | Avg. |
|---|---|---|---|---|---|---|
| GA | 2.04 | 3.82 | 1.36 | 1.08 | 2.36 | 2.13 |
| RM | 18.88 | 19.52 | 22.24 | 14.24 | 2.40 | 19.78 |

The results show that for the same test data random testing requires 9 times more timing than GA.

## 5. CONCLUSION

This paper has provided an overview of the Search-Based Software Engineering and the Search-Based Software Testing used in test data generation. The main goal is to make a study of the use of search-based optimization techniques to automate the evolution of solutions for software engineering problems. For example, real world problems such as optimizing software resource allocation, triangle classification, software clustering, component selection and prioritization for next release. Experiments show that for simple programs both work fine; there is no significant difference, but as the complexity of the program or the complexity of input domain grows, GA-based testing system significantly outperforms Random testing.

In this paper, we considered the time required for test data generation and the percentage of branch coverage; other parameters can also be considered for future work. Test data has been generated in numerals; similarly for character, string, arrays and pointers can also be tried with the genetic algorithm.

## 6. REFERENCES

[1] Mark Harman, *"The Current State and Future of SBSE"*, Future of Software Engineering (FOSE'07), IEEE Computer Society, 2007, pp. 1-16.

[2] Phil McMinn, *"Search-Based Software Test Data Generation: A Survey"*, Ph.D. Thesis, Software Testing, Verification and Reliability, 2004.

[3] Maha Alzabidi, Ajay Kumar, and A.D. Shaligram, *"Automatic Software Structural Testing by Using Evolutionary Algorithms for Test Data Generations"*, IJCSNS International Journal of Computer Science and Network Security, VOL.9, No.4, April 2009.

[4] Kamran Ghani and John A. Clark, *"Automatic Test Data Generation for Multiple Condition and MCDC Coverage"*, 2009 Fourth International Conference on Software Engineering Advances, 2009, pp. 152-157.

[5] Mark Harman, S. Afshin Mansouri and Yuanyuan Zhang, *"Search Based Software Engineering: A Comprehensive Analysis and Review of Trends, Techniques and Applications"*, April 9, 2009.

[6] J. Clark, J. J. Dolado, M. Harman, R. M. Hierons, B. Jones, M. Lumkin, B. Mitchell, S. Mancoridis, K. Rees, M. Roper, and M. Shepperd, *"Reformulating Software Engineering as a Search Problem,"* IEE Proceedings - Software, vol. 150, no. 3, 2003, pp. 161–175.

[7] André Baresel, Harmen Sthamer and Michael Schmidt, *"Fitness Function Design To Improve Evolutionary Structural Testing,"* Proceedings Of The Genetic And Evolutionary Computation Conference, 2002.

[8]   Mark Harman, *"Automated Test Data Generation using Search Based Software Engineering"*, Second International Workshop on Automation of Software Test (AST'07), IEEE Computer Society, 2007, pp. 1-2.

[9]   M. Harman and B.F. Jones, *"Search Based Software Engineering"*, Information and Software Technology, Dec. 2001, 43(41): pp. 833-839.

[10]  Praveen Ranjan Srivastava and Tai-hoon Kim, *"Application of Genetic Algorithm in Software Testing"*, International Journal of Software Engineering and Its Applications, Vol.3, No.4, Oct'2009, pp. 87-95.

[11]  Mitchell B.S., *"A Heuristic Search Approach to Solving the Software Clustering Problem"*, PhD Thesis, Drexel University, Philadelphia, PA, Jan'2002.

[12]  N.J. Tracey, *"A search-Based Automated Test-data Generation Framework for Safety-Critical Systems"*, DPhil University of York, 2000.

[13]  Y. Zhan, John A. Clark, *"A Search-Based Framework for Automatic Testing of MATLAB/Simulink Models"*, The Journal of Systems and Software 81 (2008), pp. 262-285.

[14]  P. McMinn, M. Harman, D. Binkley and Paolo Tonella, *"The Species per Path Approach to Search-Based Test Data Generation"*, International Symposium on Software Testing and Analysis (ISSTA'06), July 17-20, USA, 2006, pp. 1-11.

[15]  Yuanyuan Zhang, *"Multi-Objective Search - based Requirements Selection and Optimisation"*, Ph.D Thesis, King's College, University of London, February 2010, pp. 1-276.

## Authors

**Mrs. P. Maragathavalli** received her B.E. degree in Computer Science and Engineering from Bharathidasan University, Trichirappalli in 1998 and M.Tech. degree in Distributed Computing Systems from Pondicherry University, in 2005. She joined Pondicherry Engineering College in 2006 and currently working as Assistant Professor in the Department of Information Technology. Now she is pursuing her PhD degree in Computer Science and Engineering.
She has published research papers in International and National Conferences. She is a Life member of Indian Society for Technical Education.

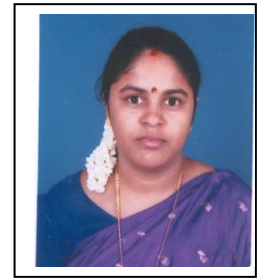